\documentclass{revtex4}
\usepackage{graphicx,amsmath,amssymb,mathrsfs}
\usepackage{epsfig}

\begin{document}
\title{Superstatistics of Blaschke products}
\author{Chris Penrose and Christian Beck}
\affiliation{School of Mathematical Sciences, Queen Mary, University
of London, London E1 4NS, UK}
\date{January 2016}

\begin{abstract}
We consider a dynamics generated by families of maps whose invariant density
depends on a parameter $a$ and where $a$ itself obeys a stochastic or periodic dynamics.
For slowly varying $a$ the long-term behavior of iterates is
described by a suitable superposition of local invariant densities.
We provide rigorous error estimates how good this approximation is.
Our method generalizes
the concept of superstatistics,
a useful technique in nonequilibrium statistical mechanics, to maps.
Our main example are Blaschke products, for which we provide rigorous
error estimate on the
difference between Birkhoff density and the superstatistical approximation.
\end{abstract}

\maketitle

\section{Introduction}
Dynamics often takes place in a changing environment. This means, given
some control parameter $a$ and a local dynamics $x_{n+1}= f_a(x_n)$ generated by some
mapping $f_a$, the control parameter $a$ itself will also slowly change in time.
In nonequilibrium statistical mechanics, these environmental
fluctuations, if taking place on a large time scale, are modeled by a
very useful concept, so-called superstatistics.
The superstatistics concept was introduced
in \cite{beck-cohen} and has since then provided a powerful tool
to describe a large variety of complex systems
for which there is change of environmental conditions
\cite{swinney,touchette,souza,chavanis,jizba,
frank,celia,straeten}.
The basic idea is to characterize the
complex system under consideration by a superposition of several
statistics,
one corresponding to local equilibrium statistical mechanics (on a mesoscopic level modeled by a
linear Langevin equation leading to locally Gaussian behavior) and the other one corresponding to a slowly
varying parameter $a$ of the system.
Essential for this approach is the fact that there is sufficient time scale separation,
i.e. the local relaxation time of the system must be much shorter than the
typical time scale on which the parameter $a$ changes.

In most applications in nonequilibrium statistical mechanics
the varying control parameter $a$ is the local inverse temperature $\beta$ of the system, i.e. $a=\beta$.
However, in some applications beyond the immediate scope of statistical mechanics
the control parameter $a$ can also have a different meaning.
There are many interesting applications
of the superstatistics concept to real-world problems,
for example to
train delay statistics\cite{briggs},
hydrodynamic turbulence \cite{prl} and cancer survival statistics
\cite{chen}. Further applications are described
in
\cite{daniels,maya,reynolds,abul-magd,rapisarda,cosmic}.

In this paper we want to extend the superstatistics concept to maps,
which usually have invariant densities different from Gaussian
distributions, and thus analyze this
problem in a more general context. Our generalization
assumes that the local dynamics is not anymore restricted to a
linear Langevin dynamics (as in the nonequilibrium statistical
mechanics applications), but given by an {\em a priori} arbitrary map with
strong mixing properties.
We will make the superstatistics concept mathematically rigorous by considering simple model
examples of local maps where everything can be proved explicitly and by estimating
the error terms.
In our new approach described here, we allow for {\em a priori} arbitrary
local invariant densities $\rho_a(x)$ and consider a
dynamics given by long term iteration of a map $f_a$ with
slowly varying $a$. This leads to a mixing of various invariant densities
$\rho_a(x)$ with different parameter $a$, in a way that we will
analyze in detail in this paper.

If $a$ changes on a long time scale, long as compared to the
relaxation time of the local map $f_a$,
the resulting long-term probability distribution of iterates
is closely approximated by a superposition of local invariant
densities $\rho_a(x)$.
For particular examples, Blaschke products, we will indeed
provide estimates of the error terms involved
and prove how fast the Birkhoff density approaches the
superstatistical approximation.
On the other hand,
if $a$ changes rapidly, then a different dynamics arises which
is not properly described by a mixing of the various
local invariant densities. Rather, in this case
one has to look at fixed points of the Perron-Frobenius operator of higher iterates of
composed maps $f_a$ with varying $a$. Depending on the time scale of the changes of $a$,
there are transition scenarios between both cases.

This paper is organized as follows. In section 2, we will introduce the superstatistics
concept for maps. We will study several examples in this section to
illustrate the concept and to motivate
our rigorous treatment in the later sections. In section 3 we state our main result,
estimating the error terms for alternating
block iteration of Blaschke products.
In section 4 we present a rigorous theory for
Blaschke products. We will prove the existence of
invariant measures, determine the invariant measure explicitly as a function
of the parameters involved
and prove our main result, an error estimate on the difference between
Birkhoff density and the superstatistical approximation.

\section{Superstatistical dynamics of maps}

Let us consider families of maps $f_a$ depending on a control parameter
$a$. These can be {\em a priori} arbitrary maps in arbitrary dimensions. Later we will
restrict ourselves to mixing maps and assume that an absolutely continuous invariant density
$\rho_a(x)$ exists for each value of the control parameter $a$.
The local dynamics is
\begin{equation}
x_{n+1}=f_a(x_n).
\end{equation}
We now allow for a time dependence of $a$ and study the long-term
behavior of iterates given by
\begin{equation}
x_n=f_{a_n}\circ f_{a_{n-1}} \circ \ldots f_{a_1} (x_0). \label{it}
\end{equation}
Clearly, the problem now requires the specification of the
sequence of control parameters $a_1, \ldots , a_n$ as well,
at least in a statistical sense. One possibility is a periodic
orbit of control parameters of length $L$. Another possibility
is to regard the $a_j$ as random variables and to specify the
properties of the corresponding stochastic process in parameter space.

In general, rapidly fluctuating parameters $a_j$ will lead to a very
complicated dynamics. However, there is a significant simplification
if the parameters $a_j$ change slowly. This is the analogue of
the slowly varying temperature parameters in the superstatistical
treatment of
nonequilibrium statistical mechanics \cite{beck-cohen, abc}.
The basic assumption of superstatistics is that an environmental control
parameter $a$ changes only very slowly, much slower than
the local relaxation time of the dynamics. For maps this means that significant changes
of $a$ occur only over a large number $T$ of iterations.
In practice, one can model this superstatistical case as follows:
One keeps $a_1$ constant for $T$ iterations ($T>>1$), then switches
after $T$ iterations to a new value $a_2$, after $T$ iterations
one switches to the next values $a_3$, and so on.

One of the simplest examples is a period-2
orbit in the parameter space. That is, we have an alternating sequence
$a_1,a_2$ that repeats itself, with switching between the two possible values
taking place after $T$ iterations. We are interested in the long-term
behavior of iterates obtained for $n\to \infty$. Possible
sequences of parameters $a_1,a_2, \ldots , a_L$ of period length $L$ could
be studied equally well, with a switching to the new parameter value always taking place after
$T>>1$ iterations. Another possibility are stochastic parameter
changes on the long time scale $T$.

To illustrate and motivate the superstatistics concept
for maps, we will now deal with three
important examples of families of maps $f_a$.

{\bf Example 1} We take for $f_a$ the asymmetric tent map
on $[0,1]$ given by
\begin{equation}
f_a(x)= \left\{
\begin{array}{ll}
\frac{1}{a} x & x \leq a \\
1-\frac{x-a}{1-a}  & x >a
\end{array}
\right. \label{atent}
\end{equation}
with $a\in (0,1)$. This example is somewhat trivial, because
the invariant density $\rho_a(x)$ is independent of $a$ and given by
the uniform distribution for any value of $a$. Hence, whatever the statistics of
the varying parameter sequence $a_1,a_2, \ldots$ is, we get for the long-term distribution of
iterates given by (\ref{it}), (\ref{atent}) the uniform distribution
\begin{equation}
p(x)=1
\end{equation}

{\bf Example 2} We take for $f_a$ a map of
linear Langevin type \cite{physica-a,dynala}.
This means $f_a$ is a 2-dimensional map given by a skew product of the form
\begin{eqnarray}
x_{n+1}&=&g(x_n) \label{ggg} \\
y_{n+1}&=&e^{-a\tau}y_{n}+\tau^{1/2}(x_n-\bar{g})
\end{eqnarray}
Here $\bar{g}$ denote the average of iterates of $g$.
It has been shown in \cite{physica-a} that for $\tau \to 0$, $t=n\tau$ finite
this deterministic chaotic map generates a dynamics equivalent to a linear
Langevin equation \cite{vKa}, provided the map $g$ has
the so-called $\varphi$-mixing property \cite{Bi}, and regarding the initial
values $x_0\in[0,1]$ as a smoothly
distributed random variable. Consequently, in this limit the
variable $y_n$ converges to the Ornstein-Uhlenbeck
process \cite{vKa}
and its stationary density is given by
\begin{equation}
\rho_\beta(y) =\sqrt{\frac{\beta}{2\pi}} e^{-\frac{1}{2}\beta y^2}
\end{equation}
The variance parameter $\beta$ of this Gaussian depends on the map $g$ and
the damping constant $a$.
If the parameter $a$
changes on a very large time scale, much larger than
the local relaxation time to equilibrium, one expects
for the long-term distribution of iterates a
mixture of Gaussian distributions with different variances $\beta^{-1}$.
For example, a period 2 orbit of parameter changes yields
a mixture of two Gaussians
\begin{equation}
p(y)=\frac{1}{2}\left( \sqrt{\frac{\beta_1}{2\pi}} e^{-\frac{1}{2}\beta_1 y^2}+\sqrt{\frac{\beta_2}{2\pi}}
e^{-\frac{1}{2}\beta_2y^2} \right) .\label{two-gau}
\end{equation}
Generally, for more complicated parameter changes
on the long time scale $T$,
the long-term distribution
of iterates $y_n$ will be  mixture of Gaussians
with a suitable weight function $h (\beta)$ for $\tau \to 0$:
\begin{equation}
p(y)\sim \int d\beta \; h(\beta)e^{-\frac{1}{2}\beta y^2}
\end{equation}
This is just the usual form of superstatistics used in
statistical mechanics, based on a mixture of Gaussians with
fluctuating variance with a given weight function \cite{beck-cohen}.
Thus for this example of skew products
the superstatistics of the map $f_a$ reproduces the concept
of superstatistics in nonequilibrium statitistical mechanics,
based on the Langevin equation. In fact, the map $f_a$ can be
regarded as a possible microscopic dynamics underlying the Langevin equation.
The random forces pushing the particle left and right are in this case
generated by deterministic chaotic map $g$ governing the dynamics of the variable
$x_n$. Generally it is possible to consider any
$\varphi$-mixing map here \cite{physica-a}.
Based on functional limit theorems, one can prove equivalence with
the Langevin equation in the limit $\tau \to 0$.



{\bf Example 3: Blaschke products}
We now want to consider further examples beyond the
immediate scope of statistical mechanics where the invariant density
of the local map
is non-Gaussian but still a full analytic treatment is possible.
Consider mappings of a complex variable $z$ given by
\begin{equation}
f(z)=b_0 \prod_{j=1}^d \frac{z-b_j}{1-\bar{b_j}z},
\end{equation}
where $|b_0| = 1$ and $|b_j| < 1$ for $j = 1, 2, \ldots, d$ and $d\geq 2$.
We are interested in a dynamics restricted to the unit circle $S^1$
and write $u\in S^1$ as
\begin{equation}
u=e^{i2\pi\varphi},
\end{equation}
so that $\varphi \in [0,1)$. According to Martin \cite{M},
and as established in the following sections in much more detail,
the invariant density of $f$ with respect to the variable $u$ is given by
\begin{equation}
\rho^u(u)= \frac{1}{2\pi} \frac{1-|z_0|^2}{|u-z_0|^2}.
\end{equation}
Here $z_0=f(z_0)$ is a fixed point of $f$.
Blaschke products usually exhibit very strong chaotic behaviour
and can be used as the map $g$ in equation (\ref{ggg})
if an extension to a Langevin dynamics is wished for for physical
reasons.
The remarkable property of Blaschke products is that the
knowledge of a fixed point $z_0$ of the map uniquely fixes the shape of
the invariant density $\rho^u$, making an analytic treatment very
convenient.

Transformation of variables $u\to \varphi$ yields the invariant density
$\rho^\varphi$ with respect to the variable $\varphi$ as
\begin{eqnarray}
\rho^\varphi (\varphi)&=& \rho^u (u) \left| \frac{du}{d\varphi} \right| \\
&=& \frac{1- |z_0|^2}{|e^{i2\pi\varphi}-z_0|^2} \\
&=&\frac{1-a^2-b^2}{1+a^2+b^2-2a\cos 2\pi \varphi -2b \sin 2\pi \varphi}.
\end{eqnarray}
Here $a$ denotes the real part and $b$ the imaginary part of the fixed point
$z_0$,
i.e. $z_0=a+ib$.

For $z_0=0$ we get $a=b=0$ and hence $\rho^\varphi (\varphi)=1$. This is
just the invariant density of a $d$-ary shift map on [0,1],
noting that for $b_j=0$
the Blaschke product becomes
\begin{equation}
z \to z^d \Longleftrightarrow \varphi \to d\cdot \varphi \; mod \; 1
\end{equation}

Another example would be a Blaschke
product with a real fixed point $z_0=a$. In this case the
invariant density is
\begin{equation}
\rho^\varphi (\varphi)= \frac{1-a^2}{1+a^2-2a\cos 2\pi \varphi}.
\end{equation}
A particular example, taken from \cite{M}, is $b_1=b_2=\frac{1}{2}$, i.e.
\begin{equation}
f(z)= \frac{(z-\frac{1}{2})^2}{(1-\frac{1}{2}z)^2}
\end{equation}
The fixed point condition $z_0=f(z_0)$ is solved by
\begin{equation}
z_0=\frac{1}{2}(7-\sqrt{45}) \approx 0.145898...=a
\end{equation}
This is the unique fixed point in the unit disk $D=\{z| \; |z| < 1\}$.

We are now in a position to explicitly do superstatistics for Blaschke products,
since the invariant densities are known explicitly as a function of the parameters of the map.
While a rigorous treatment will be worked out in detail
in the following sections, we here just consider a simple example to illustrate
the general idea. Let us consider two different Blaschke products,
and a periodic
orbit of length 2 of the parameters. For example, we may consider an alternating
dynamics of the two maps
\begin{eqnarray}
f_1(z) &=&z^2 \\
f_2(z) &=&\frac{(z-\frac{1}{2})^2}{(1-\frac{1}{2}z)^2}.
\end{eqnarray}
If we iterate $f_2$ for a long time $T$, then iterate $f_1$ for the same long
time $T$, then switch back to $f_2$, then to $f_1$, and so on, the Birkhoff
density will be a mixture
of both invariant densities. In the $\varphi$ variable we
expect to get for $T\to \infty$ the superstatistical
result
\begin{eqnarray}
\rho_\infty (\varphi)&=& \frac{1}{2}\rho_1(\varphi)+\frac{1}{2} \rho_2(\varphi)\\
&=&\frac{1}{2}\left( 1+\frac{1-a^2}{1+a^2-2a\cos 2\pi \varphi} \right)
\end{eqnarray}
with $a\approx 0.145898...$.
This will be confirmed by our rigorous treatment in the following section.
If, on the other hand, we switch maps after each iteration step,
the result will be different. In this case we need to determine the invariant density
of the composed map $f_1\circ f_2(z)$ (or $f_2\circ f_1(z)$, depending on which map
is iterated first). The composed map is again a
Blaschke product, now with $d=4$:
\begin{equation}
f_{12}(z):= f_1\circ f_2(z)=\frac{(z-\frac{1}{2})^4}{(1-\frac{1}{2}z)^4} \label{order12}
\end{equation}
The fixed point condition
\begin{equation}
z_0=f_{12}(z_0)=\frac{(z_0-\frac{1}{2})^4}{(1-\frac{1}{2}z_0)^4}
\end{equation}
yields $z_0\approx 0.0464774...=:c$ and hence the invariant density
of $f_{12}=f_1 \circ f_2$ is given by
\begin{equation}
\rho_{12}(\varphi)=\frac{1-c^2}{1+c^2-2c\cos 2\pi \varphi}.
\end{equation}
Similarly the other composed map is also a Blaschke product with $d=4$:
\begin{equation}
f_{21}(z):= f_2\circ f_1(z)=\frac{(z^2-\frac{1}{2})^2}{(1-\frac{1}{2}z^2)^2} \label{order21}
\end{equation}
The fixed point condition
\begin{equation}
z_0=f_{21}(z_0)=\frac{(z_0^2-\frac{1}{2})^2}{(1-\frac{1}{2}z_0^2)^2}
\end{equation}
yields
$z_0 = c^{\frac{1}{2}}\approx 0.0464774...^{\frac{1}{2}}\approx 0.215586...$
and hence the invariant density of $f_{21}=f_2 \circ f_1$ is given by
\begin{equation}
\rho_{21}(\varphi)=\frac{1-c}{1+c-2\sqrt{c} \cos 2\pi \varphi}.
\end{equation}
The Birkhoff density,
which describes the long-term distribution of iterates
independent of the phase of the periodic orbit,
 is then given by
$\frac{1}{2}\rho_{12}(\varphi)+\frac{1}{2}\rho_{21}(\varphi)$.

Fig.~1 shows the densities $\rho_{12}$
and  $\rho_{21}$ (dashed lines), the Birkhoff density
$\frac{1}{2}(\rho_{12}+\rho_{21})$ (dotted line) and
the superstatistical composition $\rho_\infty$ (solid line)
as a function of $\varphi$.
\begin{figure}
\epsfig{file=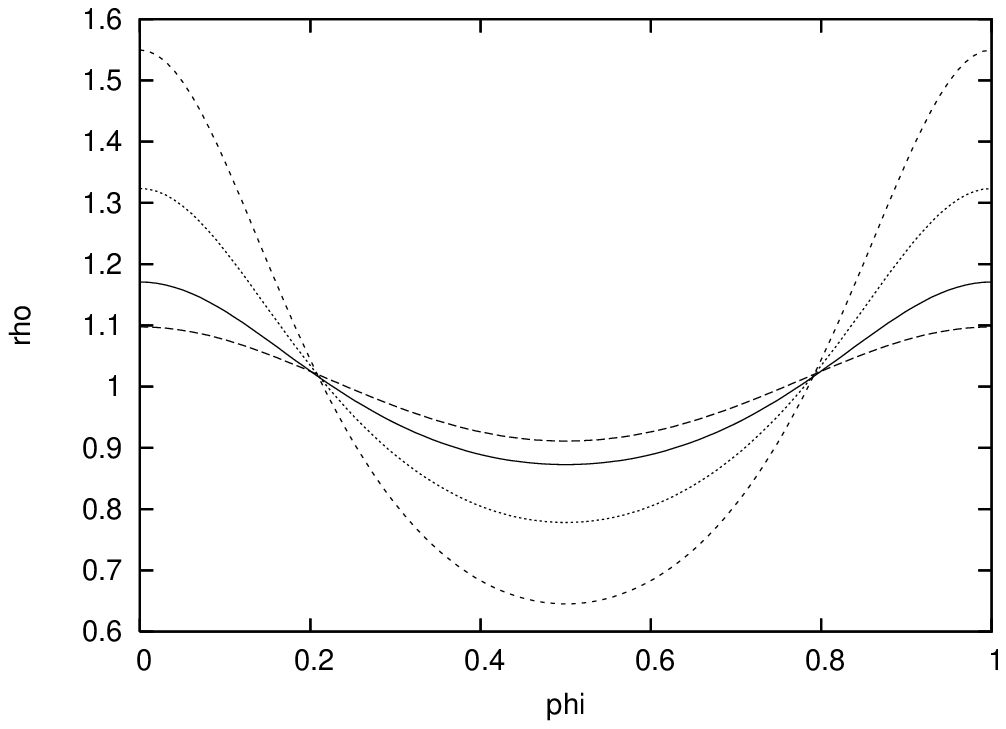, angle=270., width=10cm}
\caption{
The density $\rho_{12}(\varphi)$ (lower dashed line at $\varphi =0$)
and the density $\rho_{21}(\varphi)$ (upper dashed line at $\varphi=0$),
obtained by switching between $f_1$ and $f_2$
on the short time scale $T=1$. The Birkhoff density
$\frac{1}{2} \rho_{12} (\varphi)+ \frac{1}{2} \rho_{21}(\varphi)$
corresponds to the dotted line. In the limit $T\to \infty$, the
Birkhoff density will converge to the
superstatistical composition $\rho_\infty (\varphi)$ (solid line).}
\end{figure}
Apparently, all curves are
different.
But the difference between Birkhoff density $\frac{1}{2} \rho_{12}+
\frac{1}{2}\rho_{21}$ and the superstatistical composition $\rho_{\infty} $
will decrease if the time scale $T$ of switching between parameters
is increased.
There is a transition scenario from the Birkhoff density to the
superstatistical result $\rho_\infty$ for $T\to \infty$.

\section{Statement of main result}

Physicists use superstatistical techniques in many applications
\cite{swinney,jizba,frank,celia,straeten,briggs,daniels,reynolds,abul-magd}, but a
rigorous proof and estimate
for the error terms involved in the superstatistical approximation are missing.
It would be highly desirable to have a very general theorem delivering this.
Unfortunately, in full generality such a theorem is out of reach presently.
Hence in this paper we restrict ourselves to a first step in this direction,
a rigorous proof for particular dynamics (Blaschke products) and a
particular orbit structure in parameter space, namely blocks of two alternating
Blaschke products.

Let us prepare the mathematical background for our main theorem.
Let $\lambda_0$ be the normalized Lebesgue measure on the unit circle ${\bf S}^1$.
For any $z$ in the unit disk, let $\lambda_z$ denote the harmonic measure on
the unit circle whose density $\rho_z$ with respect to the normalized Lebesgue
measure $\lambda_0$ on ${\bf S}^1$ is given by the Poisson kernel
$\frac{1 - |z|^2}{|u - z|^2}$, $u\in {\bf S}^1$.

Suppose we are given two Blaschke products $A$ and $B$ which expand Lebesgue
measure $\lambda_0$ on the unit circle.
%
Given an arbitrary composition $C = C_l\circ\ldots\circ C_1$ (or word) with
$C_i\in\{A,B\}$ define $\lambda_C = \lambda_{\gamma}$ where $\gamma$ is the
attracting fixed point of the composition $C$. Thus
$\lambda_A = \lambda_{\alpha}$ where $\alpha$ is the attracting fixed point
of $A$ and $\lambda_B = \lambda_{\beta}$ where $\beta$ is the attracting
fixed point of $B$. We are interested in alternate block iteration of $A$
of length $m$ and of $B$ of length $n$.
When iterated cyclically the two maps $A$ and $B$ induce a Birkhoff measure
$$ \frac{1}{m+n}\{\sum_{i=1}^m\lambda_{A^i\circ B^n\circ A^{m-i}} +
	\sum_{j=1}^n\lambda_{B^j\circ A^m\circ B^{n-j}}\}\ . $$
The sum over $i$ and $j$ takes care of the fact that we average over all
possible phases of iteration of $A^m$ and $B^n$ in a cyclic way, analogous to eq.~(\ref{order12})
and (\ref{order21}) for $m=n=1$.
Our main result is that as $m$ and $n$ tend to infinity with fixed ratio $p:q$
(satisfying $p+q = 1$) this Birkhoff measure tends to the super-statistical
limit $p\lambda_A + q\lambda_B$. One thus gets in this limit a significant
simplification --- just the superstatistical approximation used by physicists.

To formulate our main theorem, it is useful to proceed to densities.
Writing $\rho_C ( = \rho_\gamma)$ for the (Poisson) density of $\lambda_C$
we obtain the quantity
$$ \frac{1}{m+n}\{\sum_{i=1}^m\rho_{A^i\circ B^n\circ A^{m-i}} +
	\sum_{j=1}^n\rho_{B^j\circ A^m\circ B^{n-j}}\} $$
as the Birkhoff density for the cycle of maps $A$ and $B$.

\medskip
\noindent
Given a Blaschke product $B$ and a point $\alpha$ in the unit
disk, define the error $\varepsilon_B(\alpha)$ by
$$ \varepsilon_B(\alpha) = \sum_{j=1}^{\infty} \varepsilon_{B^j,\alpha}\ . $$
Similarly, given a Blaschke product $A$ and a point $\beta$ in the unit
disk, define the error $\varepsilon_A(\beta)$ by
$$ \varepsilon_A(\beta) = \sum_{i=1}^{\infty} \varepsilon_{A^i,\beta}\ . $$

We are interested in the case $\alpha$ is the attracting fixed point of $A$
and $\beta$ is the attracting fixed point of $B$.

The individual terms $\varepsilon_{B^j,\alpha}$ are given by the difference
in Poisson densities $\rho_{B^j(\alpha)} - \rho_\beta$ and
the individual terms $\varepsilon_{A^i,\beta}$ are given by the difference
in Poisson densities $\rho_{A^i(\beta)} - \rho_\alpha$. These
Poisson differences converge to zero, in the supremum norm on densities,
exponentially fast as $i$ and $j$ tend to infinity.
Hence the above errors are well-defined.


\bigskip
Given $|b|$ with $0 < |b| < 1/\sqrt{3}$ put
$$ r_{|b|} = \frac{1 - |b|^2 - \sqrt{(1 + |b|^2)(1 - 3 |b|^2)}}{2 |b|^2} $$
which satisfies $0 < r_{|b|} < 1$.

\medskip
\noindent
THEOREM: Given two Blaschke products $A$ and $B$ with opposite zeroes:
$A(z) = a_0\frac{z^2 - a^2}{1 - \bar{a}^2 z^2}$ and
$B(z) = b_0\frac{z^2 - b^2}{1 - \bar{b}^2 z^2}$ (with $|a_0| = |b_0| = 1$)
and satisfying $|a|, |b| < 1/\sqrt{3}$, and given $r$ with
$\max\{r_{|a|}, r_{|b|}\} \le r < 1$ then, putting $K = \frac{2r}{1+r^2}$
($ < 1$), we have, for all $m, n\in {\bf N}$:

$$ \left| \sum_{i=1}^m \rho_{A^i\circ B^n\circ A^{m-i}} - m\rho_\alpha -
	\varepsilon_A(\beta) \right| < \frac{4r}{(1-r)^2}
	\left( \frac{K^{n+1} + K^{m+1}}{1 - K}\right)\ , $$
$$ \left| \sum_{j=1}^n \rho_{B^j\circ A^m\circ B^{n-j}} - n\rho_\beta -
	\varepsilon_B(\alpha) \right| < \frac{4r}{(1-r)^2}
	\left( \frac{K^{m+1} + K^{n+1}}{1 - K}\right)\ . $$

\medskip
\noindent
COROLLARY:
$$ \sum_{i=1}^m \rho_{A^i\circ B^n\circ A^{m-i}} +
	\sum_{j=1}^n \rho_{B^j\circ A^m\circ B^{n-j}}
	- (m\rho_{\alpha} + n\rho_{\beta}) \ \rightarrow\
	\varepsilon_A(\beta) + \varepsilon_B(\alpha) $$
exponentially fast as $m$ and $n$ tend to infinity.

\medskip
\noindent
COROLLARY: The Birkhoff density
$$ \frac{1}{m+n}\{\sum_{i=1}^m\rho_{A^i\circ B^n\circ A^{m-i}} +
	\sum_{j=1}^n\rho_{B^j\circ A^m\circ B^{n-j}}\} $$
tends to the super-statistical limit $p\rho_A + q\rho_B$ as $m$ and $n$
tend to infinity with fixed ratio $p:q$ (satisfying $p + q = 1$).

Note that the linear combination $p\rho_A +q\rho_B$ occuring above is just the analogue of the superstatistical
approximation used by
physicists, with the invariant densities
of the two different Blaschke products replacing the two different stationary
Gaussian distributions that occur
in equation (\ref{two-gau}). The limit $m\to \infty$ and $n \to \infty$ corresponds to
the assumption of time scale separation made by physicists: The system has enough
time to relax to the stationary state of $A$ before the next parameter change
takes place, changing the dynamics to $B$. The fact that $p$ and $q$ occur simply means that
it is relevant how long the system stays in state $A$ as compared to state $B$.
All this makes physical sense.

\bigskip

In the following sections we show how to arrive at this rigorous result,
by proving
statements for the existence and uniqueness
of invariant measures of Blaschke products
and estimating the relevant error terms of the superstatistical
approximation.

\section{Detailed calculations for Blaschke products}

\bigskip
\noindent
{\bf Degree $d$ maps of the circle}

\medskip
Let $\lambda_0$ be normalized Lebesgue measure on the unit circle ${\bf S}^1$.
Suppose $\tau: {\bf S}^1\rightarrow {\bf S}^1$ is $C^1$ and has degree $d > 1$.
Then its pushforward action $\tau^{\ast}: \mu\mapsto \mu\circ\tau^{-1}$ on
probability measures $\mu$ which are absolutely continuous with respect to
Lebesgue measure is given by the transfer operator ${\cal L}_{\tau}$ on
densities. Thus
$\tau^{\ast}: \rho \lambda_0\mapsto {\cal L}_{\tau}(\rho) \lambda_0$ where
$$ ({\cal L}_{\tau}(\rho))(e^{i\theta}) = \sum_{e^{i\zeta}\in\tau^{-1}
	(e^{i\theta})} \frac{\rho(e^{i\zeta})}{|\tau'(e^{i\zeta})|}\ . $$

\medskip
The main example which we are concerned with is when $\tau$ is the
restriction to ${\bf S}^1$ of a Blaschke product
$$ B(z) = b_0 \prod_{j=1}^d\frac{z - b_j}{1 - \bar{b}_j z} $$
(where $|b_0| = 1$ and $|b_j| < 1$ for $j = 1, 2, \ldots, d$).

\bigskip
\noindent
{\bf Action of Blaschke products on Poisson measures}

\medskip
For any $z$ in the unit disk, let $\lambda_z$ denote the harmonic measure on
the unit circle whose density $\rho_z$ with respect to the normalized Lebesgue
measure $\lambda_0$ on ${\bf S}^1$ is given by the Poisson kernel
$\frac{1 - |z|^2}{|u - z|^2}$, $u\in {\bf S}^1$.

\medskip
As observed in \cite{M}, a Blaschke product pushes forward Poisson measures
to Poisson measures.

\medskip
\noindent
PROPOSITION 1: If $f: {\bf D}\rightarrow {\bf D}$ is an analytic function on
the unit disk ${\bf D}$ whose extension to $\bar{\bf D}$ is continuous and
where the restriction $\tau$ to ${\bf S}^1$ takes values in ${\bf S}^1$.
Then, for all $z\in {\bf D}$, we have
$$ \tau^{\ast} \lambda_z = \lambda_{f(z)}\ . $$

\medskip
\noindent
{\it Proof.} The proof is given in \cite{M}, but we repeat it here in order
to make the paper self-consistent. Given a continuous function $\psi: {\bf S}^1\rightarrow {\bf R}$
the unique extension $\bar{\psi}$ of $\psi$ which is continuous on
$\bar{\bf D}$ and harmonic in ${\bf D}$ is given by
$$ \bar{\psi}(z) = \int_{{\bf S}^1} \psi d\lambda_z $$
where $\lambda_z$ is the harmonic measure on ${\bf S}^1$ determined by
$z\in {\bf D}$. (See \cite{Co} Chapter 10.)

Given $f: {\bf D}\rightarrow {\bf D}$ an analytic function on the unit disk
${\bf D}$ whose boundary value $\tau$ to ${\bf S}^1$ is continuous,
mapping ${\bf S}^1$ to ${\bf S}^1$, then for any continuous
$\psi: {\bf S}^1\rightarrow {\bf R}$ we have that $\bar{\psi}\circ f$ is a
harmonic function with boundary value $\psi\circ\tau$ whence equals
$\overline{\psi\circ\tau}$ (by the uniqueness theorem for harmonic extensions).

Write $\tau^{\ast}$ for the action on probability measures on ${\bf S}^1$
induced by $\tau$. Then for all $\psi: {\bf S}^1\rightarrow {\bf R}$
continuous we have
$$ \int_{{\bf S}^1} \psi d(\tau^{\ast}\lambda_z) = \frac{1}{2\pi} \int_0^{2\pi}
	\psi(e^{i\theta}) ({\cal L}_{\tau}(\rho_z))(e^{i\theta}) d\theta $$
$$ = \frac{1}{2\pi} \int_0^{2\pi} \psi(e^{i\theta}) \sum_{e^{i\zeta}\in
	\tau^{-1}(e^{i\theta})} \frac{\rho_z(e^{i\zeta})}{|\tau'(e^{i\zeta})|}
	d\theta $$
$$ = \frac{1}{2\pi} \int_0^{2\pi} \psi(\tau(e^{i\zeta})) \rho_z(e^{i\zeta})
	d\zeta $$
$$ = \int_0^{2\pi} (\psi\circ \tau)(e^{i\zeta}) d\lambda_z(\zeta) $$
$$ = \int_{{\bf S}^1} (\psi\circ\tau) d\lambda_z =
	\overline{\psi\circ\tau}(z) $$
$$ = (\bar{\psi}\circ f)(z) = \bar{\psi}(f(z)) = \int_{{\bf S}^1} \psi
	d \lambda_{f(z)} \ . $$
It follows that $\tau^{\ast} \lambda_z = \lambda_{f(z)}$ by uniqueness of the
probability measure. $\Box$

\medskip
Consequently, if $B$ is a Blaschke product with a fixed point $z$ in the
unit disk then $\lambda_z$ is an invariant measure for the action ($\tau$) on
${\bf S}^1$. With respect to such smooth invariant measure $\tilde{\lambda}$
the transfer equation ${\cal L}_{\tilde{\tau}}(\tilde{\rho}) = \tilde{\rho}$
(with $\tilde{\rho} = 1$ and $\tilde{\tau}$ the straightened circle map)
together with an upper bound on $|\tilde{\tau}'|$ shows
(via $|\tilde{\tau}'| > 1$) that $\tau$ expands such measure.
By Walters \cite{W} (Theorem 18)
there is only one invariant probability measure absolutely continuous
with respect to Lebesgue and so this must be $\lambda_z$.
The measure is mixing whence ergodic.

\medskip
If a Blaschke product of degree $d$ expands some measure on the unit circle
then there must be exactly $d - 1$ fixed points on ${\bf S}^1$, but since
there are $d + 1$ fixed points on the Riemann sphere, there must be two
further fixed points, one in each component of the complement of ${\bf S}^1$,
and both attracting (by the Schwarz lemma).

\bigskip
\noindent
{\bf Blaschke products which expand Lebesgue measure}

\medskip
Martin \cite{M} gives a sufficient condition for a Blaschke product
$$ B(z) = b_0 \prod_{j=1}^d\frac{z - b_j}{1 - \bar{b}_j z} $$
(where $|b_0| = 1$ and $|b_j| < 1$ for $j = 1, 2, \ldots, d$ and $d \ge 2$)
to expand Lebesgue measure on the unit circle. The sufficient condition is
$$ \sum_{j=1}^d\frac{1 - |b_j|}{1 + |b_j|} > 1\ . $$
(In the special case $d = 2$ and $b_1 + b_2 = 0$ this can be improved to
$$ \sum_{j=1}^2\frac{1 - |b_j|^2}{1 + |b_j|^2} > 1\ , $$
which reduces to $|b_j| < 1/\sqrt{3}$.)

\bigskip
\noindent
{\bf Dynamics on the unit disk}

\medskip
Any holomorphic self map of unit disk does not increase the {\it pseudo
hyperbolic distance}
$$ d(z,w) = \frac{|z - w|}{|1 - \bar{w}z|} $$
(a metric which is invariant under M\"{o}bius transformations preserving the
disk). In the case the self map is a Blaschke product of degree at least two
we get contraction uniform on compact subsets of the disk.

\medskip
\noindent
LEMMA 2: For $z$ and $w$ in the unit disk we have
$$ \frac{|z - w|}{|1 - \bar{w}z|} \le \frac{|z| + |w|}{1 + |z||w|}\ . $$

\medskip
\noindent
{\it Proof.}
We consider $w$ fixed and $z$ varying around a circle centre the origin and
radius $r < 1$. Then the pseudo hyperbolic distance between $z$ and $w$
is the absolute value of
$$ v = \frac{z - w}{1 - \bar{w}z}\ . $$
Inverting this yields
$$ z = \frac{v + w}{1 + \bar{w}v} $$
and the absolute value of this equals $r$. Hence
$$ |v + w|^2 = r^2 |1 + \bar{w}v|^2 $$
This can be written as
$$ \left|(1 - r^2 w\bar{w})v + (1 - r^2)w\right| = r (1 - w\bar{w}) $$
which is the equation of a circle in $v$. The centre is
$$ -\frac{(1 - r^2)w}{1 - r^2 w\bar{w}} $$
and the radius is
$$\frac{r(1 - w\bar{w})}{1 - r^2 w\bar{w}} $$
whence the maximum of $|v|$ on the circle is
$$ \frac{(1 - r^2)|w| + r(1 - |w|^2)}{1 - r^2 |w|^2} $$
$$ = \frac{|w| - r^2 |w| + r - r |w|^2}{(1 - r |w|)(1 + r |w|)} $$
$$ = \frac{(1 - r |w|)(r + |w|)}{(1 - r |w|)(1 + r |w|)}
	= \frac{r + |w|}{1 + r |w|}\ . \ \Box $$

\medskip
\noindent
PROPOSITION 3:
In the case $B$ is a Blaschke product of degree two with opposite zeros
we have
$$ \frac{d(B(z),B(w))}{d(z,w)} = \left|\frac{z + w}{1 + \bar{w}z}\right|\ . $$

\medskip
\noindent
{\it Proof.}
Introduce $[z,w] = \frac{z - w}{1 - \bar{w} z}$. Then a degree two Blaschke
product can be written $B(z) = b_0 [z,b_1] [z,b_2]$. When $B$ is a Blaschke
product of degree $d$ in $z$ then the derived map
$$ B^{\triangle}(z,w) = \frac{[B(z),B(w)]}{[z,w]} $$
turns out to be a Blaschke product of degree $d-1$ in $z$ \cite{BM}.
We prove this in the case $B$ has degree two with opposite zeros $b_1$ and
$b_2$ (written $\pm b$) obtaining the precise formula.

$$ [B(z),B(w)] = \frac{B(z) - B(w)}{1 - \overline{B(w)} B(z)} $$
$$ = \frac{b_0 [z,b_1] [z,b_2] - b_0 [w,b_1] [w,b_2]}
	{1 - \bar{b}_0 [\bar{w},\bar{b}_1] [\bar{w},\bar{b}_2]
	b_0 [z,b_1] [z,b_2]} $$
$$ = \frac{b_0\left[\frac{z^2 - b^2}{1 - \bar{b}^2 z^z}
	- \frac{w^2 - b^2}{1 - \bar{b}^2 w^2}\right]}
	{1 - b_0\bar{b}_0\left(\frac{\bar{w}^2 - \bar{b}^2}{1 - b^2\bar{w}^2}
	\right) \left(\frac{z^2 - b^2}{1 - \bar{b}^2 z^2}\right)} $$
After a short calculation one obtains
$$ [B(z),B(w)] = \frac{b_0(z - w)(z + w)}{(1 - \bar{w} z)(1 + \bar{w} z)}
	\left(\frac{1 - b^2\bar{w}^2}{1 - \bar{b}^2 w^2}\right) \ . $$
Hence
$$ B^{\triangle}(z,w) = \frac{[B(z),B(w)]}{[z,w]} $$
$$ = \frac{b_0(z + w)}{(1 + \bar{w} z)}
	\left(\frac{1 - b^2\bar{w}^2}{1 - \bar{b}^2 w^2}\right) $$
is a degree one Blaschke product in $z$ whose absolute value is
$$ \left|\frac{z + w}{1 + \bar{w}z}\right|\ .\ \Box $$

\medskip
\noindent
COROLLARY 4: We have a bound on the uniform contraction on compact subsets
of the disk, which is given by
$$ \frac{d(B(z),B(w))}{d(z,w)} \le \frac{|z| + |w|}{1 + |z||w|}\ . \Box $$

\bigskip
Write $|b|$ for the common absolute value of $b_1$ and $b_2$. When
$|b| < 1/\sqrt{3}$ put
$$ r_{|b|} = \frac{1 - |b|^2 - \sqrt{(1 + |b|^2)(1 - 3 |b|^2)}}{2 |b|^2}\ . $$

\medskip
\noindent
PROPOSITION 5: Given a Blaschke product $B$ with zeros $\pm b$ satisfying
$|b| < 1/\sqrt{3}$ and an $r$ satisfying $r_{|b|} \le r < 1$ then the closed
disk $D_r$ centre $0$ radius $r$ is mapped inside itself by $B$.

\medskip
\noindent
{\it Proof.} We first show that, for $|b| < 1$ and $r < 1$, a Blaschke
product $B$ with zeros $\pm b$ maps the disk $D_r$ inside the disk $D_s$
centre $0$ radius $s$ where
$$ s = \frac{r^2 + |b|^2}{1 + |b|^2 r^2}\ . $$
Treating
$$ B(z) = b_0 \frac{z^2 - b^2}{1 - \bar{b}^2 z^2} $$
as a function of $z^2$ (and $b^2$) we can apply the lemma and obtain
$$ \frac{|z^2 - b^2|}{|1 - \bar{b}^2 z^2|}
	\le \frac{|z^2| + |b^2|}{1 + |z^2||b^2|} $$
whence the inclusion follows since $s$ is monotone increasing in $r$.

Finally the hypothesis $r\ge r_{|b|}$ then guarantees that $s\le r$. $\Box$

\medskip
If $r$ satisfies the hypotheses for two Blaschke products $A$ and $B$, each
with opposite zeros, then an iterated function system on $D_r$ with uniform
contraction $K = \frac{2r}{1 + r^2}$ (with respect to the pseudo-hyperbolic
distance on ${\bf D}$) is created. This is the situation considered by
Hutchinson \cite{H}.

When iterated cyclically the two maps $A$ and $B$ induce a Birkhoff measure
$$ \frac{1}{m+n}\{\sum_{i=1}^m\lambda_{A^i\circ B^n\circ A^{m-i}} +
	\sum_{j=1}^n\lambda_{B^j\circ A^m\circ B^{n-j}}\}\ . $$

\bigskip
\noindent
{\bf Superstatistics for Blaschke products}

\medskip
The map $z\mapsto\lambda_z$ (from the unit disk) is continuous relative to
the supremum norm on densities, as is seen in the following

\medskip
\noindent
PROPOSITION 6: For $z$ and $w$ in the unit disk and $u$ in the unit circle
$$ \left|\frac{1 - |w|^2}{|u - w|^2} - \frac{1 - |z|^2}{|u - z|^2}\right|
	\ \le\ \frac{2 |z - w|}{(1 - |z|)(1 - |w|)}\ . $$

\medskip
\noindent
{\it Proof.} For $|z| < 1$ and $|u| = 1$ we have
$$ \frac{1 - |z|^2}{|u - z|^2} = \frac{1 - |z|^2}{|(u - z)(1 - \bar{z}u)|} $$
$$ = \left|\frac{1}{u - z} + \frac{\bar{z}}{1- \bar{z}u}\right|\ . $$
Now $$ \frac{1}{u - w} - \frac{1}{u - z} = \frac{w - z}{(u-w)(u-z)} $$
and $$ \frac{\bar{w}}{1 - \bar{w}u} - \frac{\bar{z}}{1 - \bar{z}u} =
	\frac{\bar{w} - \bar{z}}{(1 - \bar{w}u)(1 - \bar{z}u)}\ . $$
It follows, by the triangle inequality, that
$$ \left|\frac{1 - |w|^2}{|u - w|^2} - \frac{1 - |z|^2}{|u - z|^2}\right|
	\le \left|\frac{w - z}{(u - w)(u - z)}\right| +
	\left|\frac{\bar{w} - \bar{z}}{(1 - \bar{w}u)(1 - \bar{z}u)}\right| $$
$$ = \frac{|w - z|}{|(u - w)(u - z)|} + \frac{|\bar{w} - \bar{z}|}
	{|(\bar{u} - \bar{w})(\bar{u} - \bar{z})|} $$
$$ = \frac{2 |w - z|}{|u - w|\cdot |u - z|} \le \frac{2 |w - z|}
	{(1 - |w|)(1 - |z|)}\ .\ \Box $$

\medskip
\noindent
DEFINITION 7: As mentioned before in section 3, given a Blaschke product
$B$ and a point $\alpha$ in the unit
disk, define the error $\varepsilon_B(\alpha)$ by
$$ \varepsilon_B(\alpha) = \sum_{j=1}^{\infty} \varepsilon_{B^j,\alpha}\ . $$
Similarly, given a Blaschke product $A$ and a point $\beta$ in the unit
disk, define the error $\varepsilon_A(\beta)$ by
$$ \varepsilon_A(\beta) = \sum_{i=1}^{\infty} \varepsilon_{A^i,\beta}\ . $$

We are interested in the case $\alpha$ is the attracting fixed point of $A$
and $\beta$ is the attracting fixed point of $B$.

The individual terms $\varepsilon_{B^j,\alpha}$ are given by the difference
in Poisson densities $\rho_{B^j(\alpha)} - \rho_\beta$ and
the individual terms $\varepsilon_{A^i,\beta}$ are given by the difference
in Poisson densities $\rho_{A^i(\beta)} - \rho_\alpha$. These
Poisson differences converge to zero, in the supremum norm on densities,
exponentially fast as $i$ and $j$ tend to infinity, by Proposition 6.
Hence the above errors are well-defined.

Given an arbitrary composition $C = C_l\circ\ldots\circ C_1$ (or word)
with $C_i\in\{A,B\}$ define $\varepsilon_C := \rho_C - \rho_{C_l}$
where $\rho_C = \rho_{\gamma}$ where $\gamma$ is the attracting fixed
point of the composition $C$. (Thus $\rho_{C_l} = \rho_{\alpha}$ if
$C_l = A$ and $\rho_{C_l} = \rho_{\beta}$ if $C_l = B$.)

\medskip
\noindent
THEOREM 8: Given two Blaschke products $A$ and $B$ with opposite zeroes:
$A(z) = a_0\frac{z^2 - a^2}{1 - \bar{a}^2 z^2}$ and
$B(z) = b_0\frac{z^2 - b^2}{1 - \bar{b}^2 z^2}$ (with $|a_0| = |b_0| = 1$)
and satisfying $|a|, |b| < 1/\sqrt{3}$, and given $r$ with
$\max\{r_{|a|}, r_{|b|}\} \le r < 1$ then, putting $K = \frac{2r}{1+r^2}$
($ < 1$), we have, for all $m, n\in {\bf N}$:

$$ \left| \sum_{i=1}^m \varepsilon_{A^i\circ B^n\circ A^{m-i}} -
	\varepsilon_A(\beta) \right| < \frac{4r}{(1-r)^2}
	\left( \frac{K^{n+1} + K^{m+1}}{1 - K}\right)\ , $$
$$ \left| \sum_{j=1}^n \varepsilon_{B^j\circ A^m\circ B^{n-j}} -
	\varepsilon_B(\alpha) \right| < \frac{4r}{(1-r)^2}
	\left( \frac{K^{m+1} + K^{n+1}}{1 - K}\right)\ . $$

\noindent
{\it Proof.} For all $i$ with $1\le i\le m$ we have
$$ \varepsilon_{A^i\circ B^n\circ A^{m-i}}=\rho_{A^i\circ B^n\circ A^{m-i}}
	- \rho_{\alpha} $$
and $\varepsilon_{A^i,\beta} = \rho_{A^i(\beta)} - \rho_{\alpha}$. Hence
$$ \varepsilon_{A^i\circ B^n\circ A^{m-i}} - \varepsilon_{A^i,\beta} =
	\rho_{A^i\circ B^n\circ A^{m-i}} - \rho_{A^i(\beta)} $$
Furthermore
$$ \left| \rho_{A^i\circ B^n\circ A^{m-i}} - \rho_{A^i(\beta)} \right|
	\le \frac{2}{(1-r)^2} |A^i(B^n(\gamma)) - A^i(\beta)| \le
	\frac{2}{(1-r)^2} \cdot 2r K^{i+n} $$
(since $d(\gamma,\beta)\le \frac{2r}{1+r^2}$) where $\gamma$ is the fixed
point of $A^m B^n$. Then, for $i > m$, we have
$$ \left| \varepsilon_{A^i,\beta} \right| \le
	\frac{2}{(1-r)^2} |A^i(\beta) - \alpha| \le
	\frac{2}{(1-r)^2} \cdot 2r K^i  $$
(since $d(\beta,\alpha)\le \frac{2r}{1+r^2}$). Summing over all $i\ge 1$ gives
the first conclusion. A similar argument gives the second conclusion. $\Box$

\medskip
\noindent
COROLLARY 9:
$$ \sum_{i=1}^m \rho_{A^i\circ B^n\circ A^{m-i}} +
	\sum_{j=1}^n \rho_{B^j\circ A^m\circ B^{n-j}}
	- (m\rho_{\alpha} + n\rho_{\beta}) \ \rightarrow\
	\varepsilon_A(\beta) + \varepsilon_B(\alpha) $$
exponentially fast as $m$ and $n$ tend to infinity.

\medskip
\noindent
COROLLARY 10: The Birkhoff density
$$ \frac{1}{m+n}\{\sum_{i=1}^m\rho_{A^i\circ B^n\circ A^{m-i}} +
	\sum_{j=1}^n\rho_{B^j\circ A^m\circ B^{n-j}}\} $$
tends to the super-statistical limit $p\rho_A + q\rho_B$ as $m$ and $n$
tend to infinity with fixed ratio $p:q$ (satisfying $p + q = 1$).


\end{document}